\documentclass{article}

\usepackage{PRIMEarxiv}

\usepackage{authblk}

\usepackage[utf8]{inputenc} 
\usepackage[T1]{fontenc}    
\usepackage{hyperref}       
\usepackage{url}            
\usepackage{booktabs}       
\usepackage{amsfonts, amsmath,amssymb}       
\usepackage{nicefrac}       
\usepackage{microtype}      
\usepackage{lipsum}
\usepackage{fancyhdr}       
\usepackage{graphicx}       
\graphicspath{{media/}}     

\title{Physics-Based Acoustic Holograms
}

\author[1]{Antonio Stanziola}
\author[1]{Ben T. Cox}
\author[1]{Bradley E. Treeby}
\author[1,2,*]{Michael D. Brown}
\affil[1]{Dept. of Medical Physics and Biomedical Engineering, University College London, Gower St, London, WC1E 6BT}
\affil[2]{Dept. of Neuroscience,\ Erasmus\ MC,\ Rotterdam,\ Netherlands}
\affil[*]{\texttt{michael.brown.13@ucl.ac.uk}}

\begin{document}
\maketitle

\begin{abstract}
Advances in additive manufacturing have enabled the realisation of inexpensive, scalable, diffractive acoustic lenses that can be used to generate complex acoustic fields via phase and/or amplitude modulation. However, the design of these holograms relies on a thin-element approximation adapted from optics which can severely limit the fidelity of the realised acoustic field. Here, we introduce physics-based acoustic holograms with a complex internal structure. The structures are designed using a differentiable acoustic model with manufacturing constraints via optimisation of the acoustic property distribution within the hologram. The holograms can be fabricated simply and inexpensively using contemporary 3D printers. Experimental measurements demonstrate a significant improvement compared to conventional thin-element holograms.
\end{abstract}


\section{Introduction}
Acoustic holograms are a low-cost technically simple alternative to phased arrays for the generation of high-fidelity, three-dimensional, distributions of acoustic pressure and phase \cite{melde2016holograms,lalonde1993field}. They are thin, 3D printable, lenses designed to map a continuous-wave incident field onto a pre-calculated phase distribution that subsequently diffracts to form a desired field. Acoustic holograms have applications across biomedical and physical acoustics, including transcranial ultrasound stimulation \cite{jimenez2019holograms}, ultrasound therapy \cite{maimbourg20183d}, compressive imaging \cite{kruizinga2017compressive}, cell assembly \cite{melde2023compact}, and particle trapping \cite{cox2019acoustic}.

To date, acoustic holograms have been direct analogues to thin optical holograms, and their application in acoustics has employed the same underlying thin-element approximation for their design. In other words, a phase and amplitude modulating thin sheet is designed that in practice is fabricated as a hologram of finite thickness comprising independent pixels (columns) of varying thickness \cite{melde2016holograms,jimenez2019holograms}. 
However, the thin-element approximation is known to increasingly break down as the hologram features approach the size of the wavelength \cite{levy2004thin}, and acoustic holograms nearly always operate in this regime. By ignoring the physical structure of the lens and the resulting complex 3D wavefield within the lens during the design, the hologram capabilities can be severely limited. Hence, there is a need for new types of holograms and hologram design methodologies that don't employ such simplifying assumptions.

Here, we introduce physics-based acoustic holograms that are designed using a differentiable acoustic model that incorporates the complex 3D wavefield within the hologram, along with manufacturing constraints, during the design process. This enables the experimental realization of holograms that can render complex acoustic fields with high-fidelity. The lenses are designed using gradient-based optimisation, which is solved in a computationally efficient way using a differentiable acoustic simulator. This allows the convenient use of automatic differentiation with respect to the lens material properties without the need to manually derive the adjoint.

\section{Lens Design}
We consider here a transducer emitting a single frequency wave propagating through a heterogeneous hologram in an otherwise homogeneous medium. The hologram is designed to control the spatially varying acoustic field outside of the lens, for example, the pattern of acoustic pressure at a target depth $d$. The internal material structure of the hologram is represented by the vector $\theta=(c, \rho, \alpha)$ which captures the heterogeneous distribution of sound speed $c$, density $\rho$, and absorption $\alpha$. For a given transducer, the resulting 3D acoustic wavefield $P$ is determined by the hologram material properties $\theta$, written here as $P(\theta)$.

\begin{figure}  
\centering
    \includegraphics[width=0.5\columnwidth]{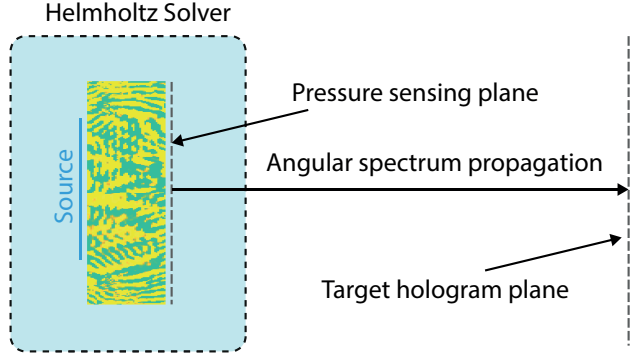}
    \caption{Setup for the acoustic hologram simulation. The Helmholtz equation is solved in a small domain surrounding the lens and source. The field is then extracted in a plane located next to the lens, opposite to the source, and propagated to the target plane using the angular-spectrum method. The wavefield at the target plane is then compared to the one of interest, and the gradients of the loss function used to iteratively update the material parameters of the lens.}
    \label{fig:simulation_setup}
\end{figure}

The wavefield at any position can be found by solving the heterogeneous Helmholtz equation \cite{stanziola2022j}. However, to improve computational efficiency, here the 3D wavefield $P(\theta)$ is solved using the Helmholtz equation in a confined spatial region that surrounds the lens and the transducer (see Fig. \ref{fig:simulation_setup}). The resulting complex acoustic field at the target depth $d$ for a given hologram, $Q(\theta)$, is then calculated by propagating the field through the surrounding homogeneous medium using the angular-spectrum method (AS) \cite{zeng2008evaluation}, denoted $\mathcal A_d$. This can be written as

\begin{equation}
Q(\theta) = \mathcal A_d \mathcal S_r P(\theta) \enspace
\label{eq:propagator}
\end{equation}

\noindent where $\mathcal S_r$ as an operator that extracts field values in a plane at distance $r$ from the lens. If the lens axis aligns with a domain axis, this operator, in discrete form, is equivalent to taking a slice of the 3D array representing $P$ at a specific index along a specific dimension. Here, the acoustic fields are solved using the j-Wave package \cite{stanziola2022j}, which uses Fourier spectral methods to solve the heterogeneous Helmholtz equation and implements the angular-spectrum method outlined in \cite{zeng2008evaluation}.

To design the lens, a real-valued loss function $\ell(Q,Q_0)$ is minimised, where $Q_0$ is the target field. If this function is differentiable with respect to its argument (e.g., mean squared error or normalized cross-correlation operator), gradient-based optimization methods can be used to adjust the values of $\theta$. As the numerical implementation of Eq.\ \eqref{eq:propagator} using j-Wave is written in a differentiable language \cite{stanziola2022j,stanziola2021jaxdf}, the chain rule can then be applied to find the gradient of the loss function with respect to $\theta$ (backpropagation is a specific instance of this). 

The optimization of the holograms is carried out with gradient descent. The loss function used depends on the required hologram. For example, for focused lenses, the acoustic power at the target location could be maximized. Here, we maximize the correlation between the amplitude of the acoustic field at the target depth $q = |Q|$ to a target amplitude $q_0 = |Q_0|$, regularized with an intensity term
\begin{equation}
    \mathbf \ell(q_0,q) = - \frac{\|q_0^*q\|}{\|q_0\|\|q\|} - \lambda \|q\| \enspace.
\end{equation}
Here $\lambda$ is the weighting for the intensity term, which is chosen experimentally, and $\|\cdot\|$ represents the $\ell_2$ norm. The regularization term helps to increase the diffraction efficiency of the designed lens by penalizing over-use of absorption and backward scattering.

\section{Lens Optimization and Fabrication}
To demonstrate the potential of physics-based acoustic holograms, holograms were designed for fabrication using a Stratasys J835 (Stratasys, Edina, MN, US). This allows for simultaneous printing of multiple materials with differing mechanical properties in a single component with a spatial resolution of $<$200 $\mu$m. These properties can vary from rubber-like Agilus30 to rigid or Poly(methyl methacrylate)-like (PMMA) veroClear \cite{bakaric2021measurement}. Note, that while the chosen fabrication method is limited to discrete materials, continuous medium properties can be approximated by sub-wavelength dithering of binary material components \cite{aghaei2022ultrasound}. 

The constraints of the chosen fabrication method were incorporated into the lens design by parameterizing the material distribution using a mixture model. The values for absorption, sound speed and density were defined as
\begin{equation}
    \theta = \left [ \begin{pmatrix}
 c_1\\ 
 \rho_1\\ 
 \alpha_1
\end{pmatrix} - \begin{pmatrix}
 c_0\\ 
 \rho_0\\ 
 \alpha_0
\end{pmatrix} \right ]\sigma(\gamma) + \begin{pmatrix}
 c_1\\ 
 \rho_1\\ 
 \alpha_1
\end{pmatrix},
\end{equation}
where $\gamma$ is a real scalar field tuned by the optimization, $\sigma(x) = e^x/(1 + e^x)$
is the sigmoid function and $(c_i, \rho_i, \alpha_i)$ are the acoustic parameters of the $i-$th material, which correspond to the two materials used for printing which have been previously characterised \cite{bakaric2021measurement}.
Incorporating the material properties into the optimisation ensures that the range of material parameters is bounded by those that are physically realizable, and also that the acoustic parameters don’t fall in a region for which the numerical simulation is unstable or inaccurate. Note that $\gamma$ is the same for all three acoustic parameters (it identifies the material mixture). 

We illustrate the performance of the proposed method experimentally. For the target field a pattern that has been previously realised was chosen: the dove of peace \cite{melde2016holograms}. The source was a 25.4 mm PZT disc (Olympus, Japan), the design frequency was 2 MHz, the lens thickness and diameter were 6 mm and 50 mm respectively, and the target depth was 12 mm after the output surface of the lens (18 mm from the source surface). We optimized the lens using the Adam optimizer \cite{kingma2014adam} with parameters $\beta = [0.9,0.9]$ and learning rate = $0.4$ for 500 iterations.

\begin{figure}
    \centering
    \includegraphics[width=0.5\columnwidth]{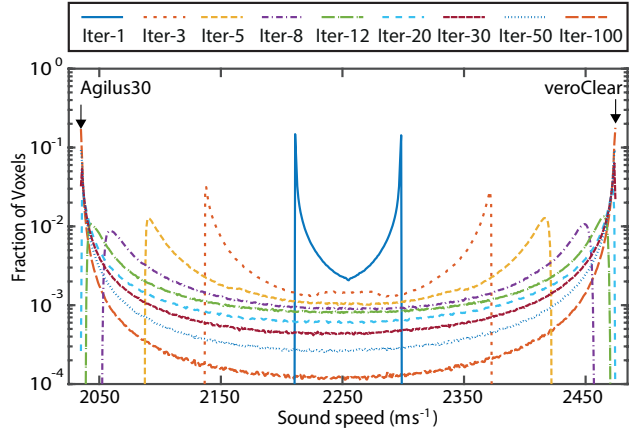}
    \caption{Evolution of sound speed distribution within the lens over multiple iterations. The distribution gets increasingly pushed to the constraints provided by the fabrication method resulting in a highly binarised lens that can be directly fabricated. This binarisation process maximises the scattering strength of the lens.}
    \label{sound_speed}
\end{figure}

The gradient descent rapidly converged to a highly binarised distribution (Fig. \ref{sound_speed}). This meant that it was not necessary to employ means for approximating continuous acoustic parameters on fabrication. The optimised lens could instead be directly fabricated after simple thresholding as the error introduced in doing so was small. For the hologram in Fig. 3(a) the difference in correlation with target field for the continuous vs binarised hologram was only 1.5$\%$ when compared using k-Wave \cite{treeby2010k}.

\begin{figure}[t!]
    \centering
    \includegraphics[width=0.6\columnwidth]{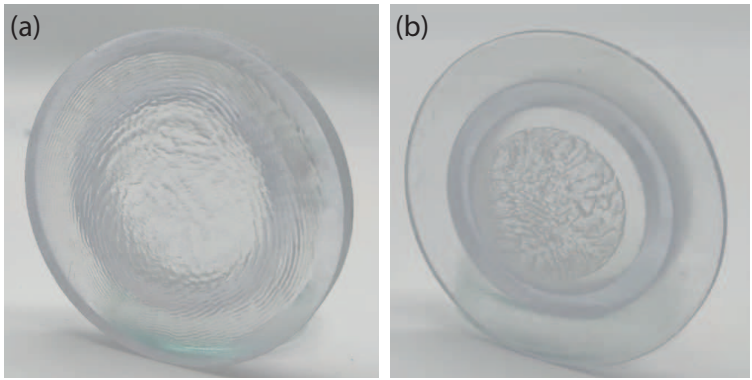}
    \caption{(a) Photograph of holographic lens designed using differentiable simulation. (b) Photograph of holographic lens designed using thin-element approximation.}
    \label{exp_photo}
\end{figure}

The hologram used for experimental comparison was taken from iteration 50 of the gradient descent. The impact of iteration choice on experimental performance is discussed in Supp.\ Sec.\ I. For comparison with the physics-based hologram, a hologram was also designed using the thin-element approximation as outlined in \cite{melde2016holograms}. This was designed for the same input parameters, however, the diameter was adjusted to 25.4 mm to match that of the piston source, since the thin-element approximation has no ability to utilise a larger aperture. Both fabricated holograms are shown in Fig.\ \ref{exp_photo}. 

The output field was characterised for both holograms in a custom-built test tank with a three-axis computer controlled positioning system using a 0.2 mm needle hydrophone (Precision Acoustics, Dorchester). The hologram was attached directly to the front surface of the transducer which was excited using a high-voltage impulse (JSR DPR300, Imaginant, NY, USA). The field was recorded over a $54\times54$ mm area with a step size of 0.3 mm. This scan was centered on the middle of the hologram, at a distance of approximately 1.2 cm from the output surface. Signals were recorded with a M4i.4421-x8 digitiser (Spectrum, Germany) using 100 averages. The acoustic field at 2 MHz was extracted at each measurement position and the planar-data was projected using AS to reconstruct the 3D acoustic field. The target depth was then identified in each dataset by maximizing the correlation with the target field. 

\begin{figure*}  
\centering
    \includegraphics[width=\columnwidth]{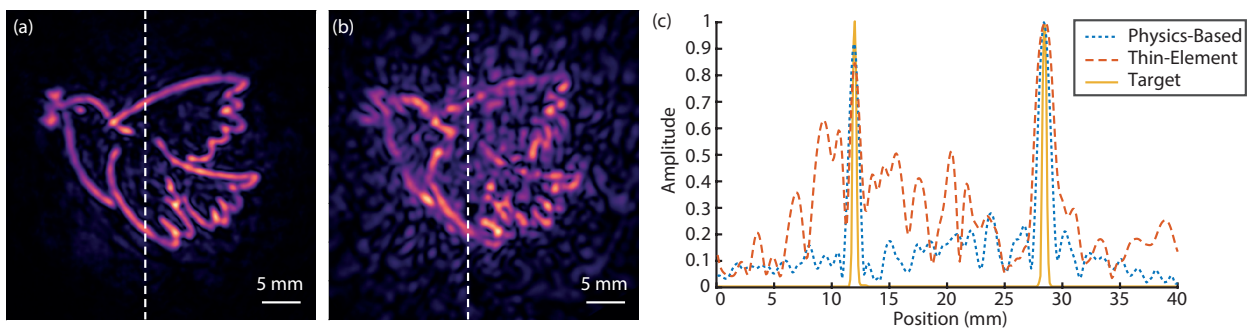}
    \caption{(a) Experimentally measured field at target depth for acoustic hologram designed using differentiation through a full-wave model. (b) Experimentally measured field at target depth for acoustic hologram designed using thin-element approximation. (c) Normalised pressure along the dashed-lines in (a) and (b) compared to target field. The lower background in the physics-based hologram is clearly apparent. The target depth for both (a) and (b) was 18 mm from the source.}
    \label{exp_data}
\end{figure*}

The acoustic field generated at the target depth by both the thin-element and physics-based holograms are shown in Fig.\ \ref{exp_data}. The field of the physics-based hologram has significantly improved fidelity in comparison to the thin-element hologram. This confirms that a more accurate model of the underlying physics can significantly enhance the performance of acoustic holograms. The contrast to noise ratio, evaluated by comparing average pressure over the target pattern with the average background pressure, for the physics based hologram is 7.6 compared with 3.8 for the thin-element hologram. This improvement is apparent when looking at normalised cross-sections through both experimental target fields compared to the target distribution (Fig.\ \ref{exp_data} (c)). The correlation coefficients with the target distribution are 0.75 and 0.56 for the physics-based and thin-element holograms, respectively. This demonstrates that, in addition to a lower background, the target distribution is also more closely approximated using a physics-based approach. 

\section{Discussion and Conclusions}
We have introduced physics-based acoustic holograms designed using a differentiable acoustic model that more accurately accounts for the complex 3D wavefield within the hologram. This approach significantly improves the fidelity of the generated acoustic fields compared to traditional thin-element approximation methods. The gradient-based optimization technique employed is computationally efficient and adaptable for various applications. This includes existing applications for acoustic holograms such as transcranial ultrasound stimulation (TUS), ultrasound therapy, and acoustic assembly, where, in addition to the enhanced fidelity that can be achieved, additional aspects of the physics can also be incorporated. For TUS, reverberations between the skull and hologram can be factored into the design (see Fig. S3), while for assembly, acoustic heterogeneity in the arrangement could be modelled. An open-source Python package, \texttt{holab - https://github.com/ucl-bug/holab}, has been released to facilitate further exploration and experimentation in hologram optimization. 

This approach is an application of inverse-design or topological optimisation which has seen broad application. In nano-photonics for example, it has been applied to the design of metasurfaces \cite{mansouree2020multifunctional} and non-linear optical switches \cite{hughes2018adjoint}. These outputs can, however, be challenging to physically realise due to fabrication constraints \cite{molesky2018inverse} and there is often still a need to simplify the physics (e.g., using periodic or axisymmetric structures or uncoupled unit-cells) \cite{hughes2021perspective}. Here, we have demonstrated an approach that accurately incorporates fabrication constraints, doesn't simplify the physics beyond linear acoustic theory, and converges to results that can be directly, and inexpensively, fabricated.  

In future, this method has the potential to be extended to use non-linear acoustic materials, potentially allowing the construction of acoustic systems that perform non-linear computations. This approach has been applied, for example, to generate optical neural networks \cite{wang2023image} that perform tasks such as image classification.

\bibliographystyle{unsrt}  
\bibliography{sample}

\cleardoublepage

\appendix

\renewcommand{\thefigure}{S\arabic{figure}}

\setcounter{figure}{0}

\section*{Physics-Based Acoustic Holograms --- Supplementary Information}

\subsection*{Gradient descent iteration and over-fitting}
The gradient descent for the example shown in Fig.\ \ref{exp_data} was observed to converge within $\sim$50 iterations to a highly binary distribution for the Adam optimiser parameters used in this work. This distribution could be fabricated by applying a simple threshold to the medium parameters and 3D printing the resulting binary structure. Running the gradient descent beyond this point produced further changes in the structure along with small reductions in the loss. 

\begin{figure*}  
\centering
    \includegraphics[width=\columnwidth]{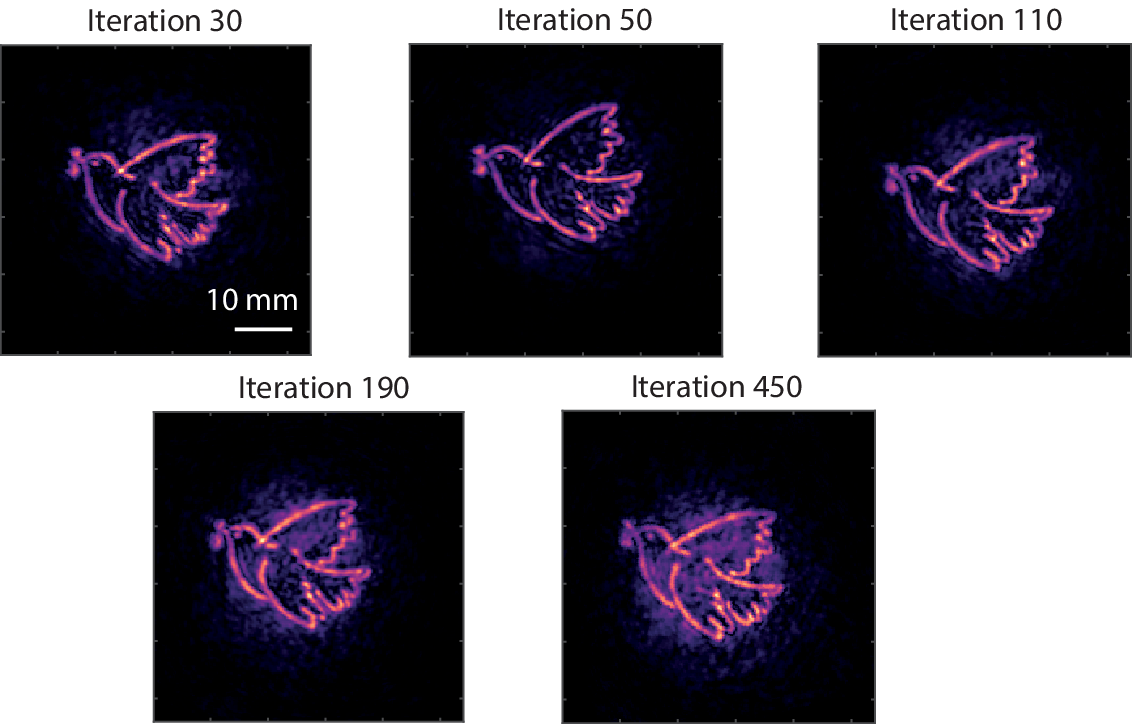}
    \caption{Experimentally measured field from 5 lenses fabricated after binarising the gradient descent output for iterations 30, 50, 110, 190, and 450.}
    \label{overfitting_figure}
\end{figure*}

To test which iteration performed best experimentally, 5 holograms were fabricated taken from iterations 30, 50, 110, 190, and 450. Each was measured experimentally with the set-up described in Sec. III with the field of each hologram shown in Fig.\ \ref{overfitting_figure}. Between iterations 30 and 50, there is an enhancement in the fidelity of the output field. This is because by iteration 30 the material distribution has not become fully binarised and an error is introduced in preparing the file for fabrication. This discrepancy could be potentially be reduced by employing sub-wavelength dithering \cite{aghaei2022ultrasound}, however, this introduces additional technical complexity. Beyond iteration 50 the fidelity of the experimentally measured fields decreases in a way that is not anticipated by the model. 

\begin{figure}  
\centering
    \includegraphics[width=0.5\columnwidth]{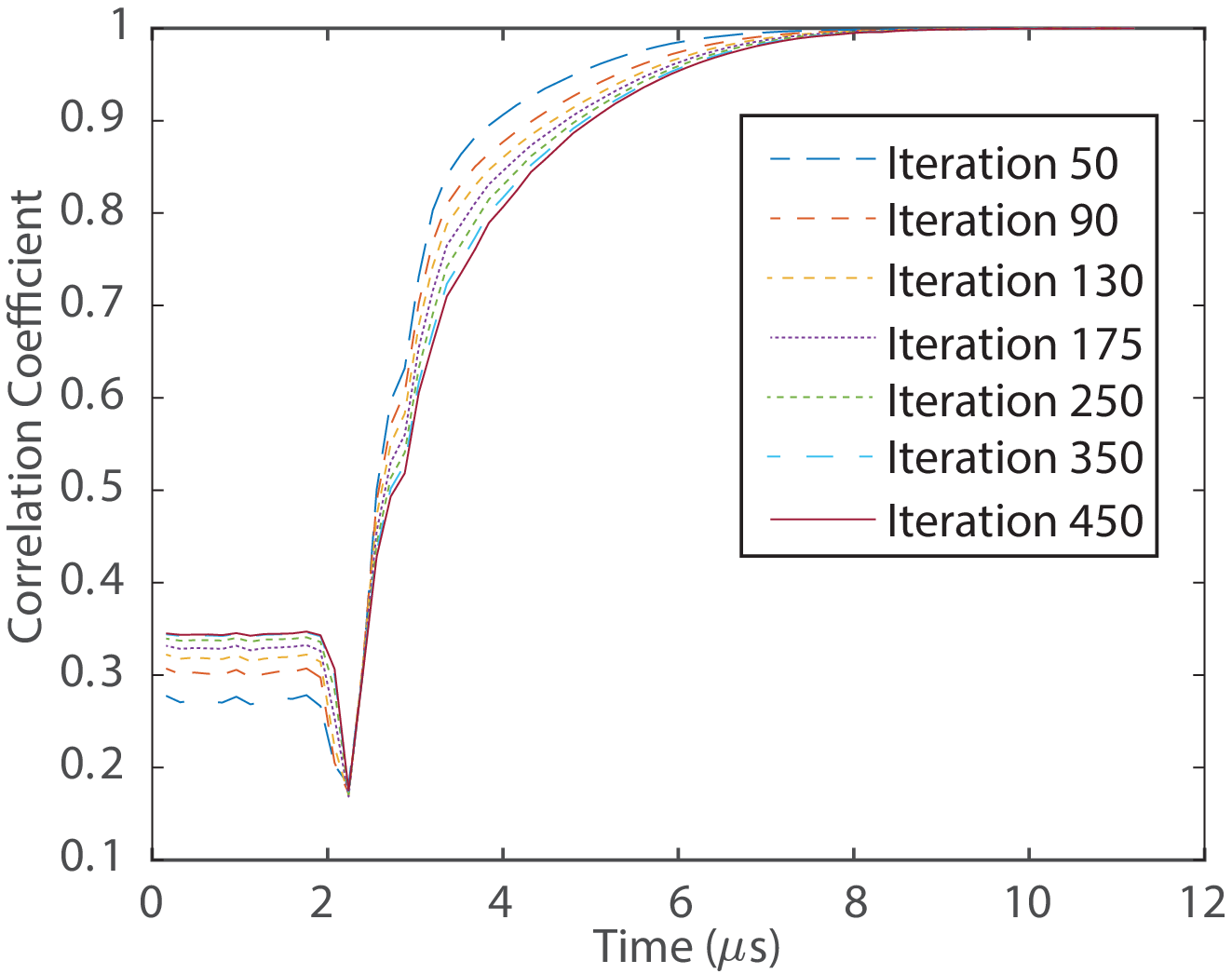}
    \caption{Variation in correlation coefficient of windowed field at 2 MHz over target-depth as a function of time for 7 different holograms evaluated using time-domain k-Wave simulations.}
    \label{correlation_figure}
\end{figure}

To investigate this discrepancy, time-domain simulations were performed through holograms from different iterations of the gradient descent using the k-Wave toolbox \cite{treeby2010k}. The field at 2 MHz directly after the hologram was extracted from each simulation using a running temporal window and propagated to the target depth using the angular spectrum method (AS). The correlation coefficient with the steady state field as a function of time for each hologram is shown in Fig.\ \ref{correlation_figure}. 

As the gradient descent iteration increases, the field more slowly converges to steady state. This suggests that the model is employing higher degrees of multiple scattering in the hologram profile. Such structures are more sensitive to numerical-experimental mismatch arising from fabrication errors hence worse experimental performance. This can be avoided by picking an early iteration for which the binarisation error on the hologram is small, as was done here. In the future, the gradient descent could be regularised to eliminate this behaviour, for example, by optimising in the time-domain and windowing the propagated field.

\subsection*{Transcranial focusing}
\begin{figure*}  
\centering
    \includegraphics[width=\columnwidth]{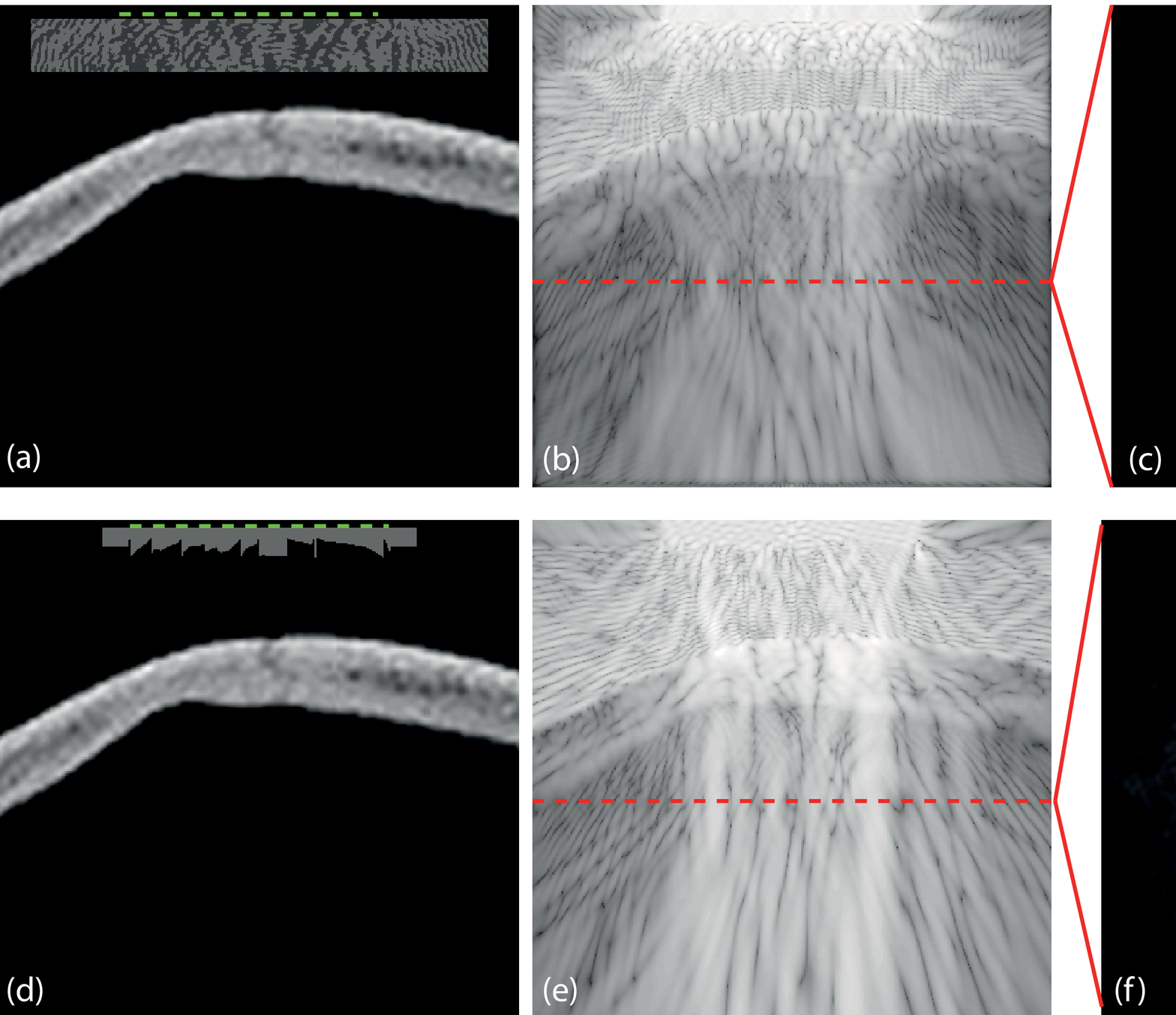}
    \caption{Numerical experiment illustrating the potential of physics-based holograms for transcranial focusing. (a) Sound speed along a cross-section through physics-based hologram and aberrator. The source is denoted by the dashed line. (b) Acoustic pressure generated through same cross-section simulated using k-Wave toolbox. The red-dashed line denotes the target depth. (c) Pressure generated over target depth by physics-based hologram. The target distribution is clearly realised. (d) Sound speed along a cross-section through thin-element hologram and aberrator. The source is denoted by the dashed line. (b) Acoustic pressure generated through same cross-section simulated using k-Wave toolbox. The red-dashed line denotes the target depth. (c) Pressure generated over target depth by thin-element hologram. The target distribution suffers from significant aberrations due to incomplete modeling of the relevant physics.}
    \label{skull_figure}
\end{figure*}

A numerical experiment was carried out to investigate the potential of physics-based holograms for transcranial focusing as shown in Fig.\ \ref{skull_figure}. The hologram was designed to form the same target used for the experiments in the main paper (a dove) behind an aberrator. The source was a 25.4 mm PZT transducer, the frequency was 2 MHz, the lens thickness and diameter were 6 and 50 mm, and the target depth was 24.7 mm from the source surface. The aberrator structure, shown in Fig.\ \ref{skull_figure} (a) and (d), was adapted from a CT scan of an ex-vivo skull. The optimisation of the hologram was performed via back-propagation through a single Helmholtz solver domain that incorporated both the hologram and the skull. This meant wave interactions both within the hologram and between the hologram and skull were fully considered. The dove is clearly realised at the desired depth behind the aberrator using the physics-based approach (Fig.\ \ref{skull_figure} (a-c)). For comparison a thin-element hologram was also designed and simulated for the same parameters. For the thin-element hologram,  back-propagation through the aberrator followed by phase-conjugation was used. The errors introduced by both the simplifying assumptions of the thin-element approximation along with the absence of reverberations between the skull and hologram from the model result in significant aberrations in the realisation of the target field (Fig.\ \ref{skull_figure} (d-f)).

\end{document}